\documentclass[%
reprint,
showpacs,
showkeys,
nofootinbib,
nobibnotes,
amsmath,
amssymb,
aps, 
superscriptaddress
]{revtex4-1}

\usepackage[english]{babel}
\usepackage[utf8x]{inputenc}
\usepackage[T1]{fontenc}
\usepackage{lineno}
\usepackage{amsmath}
\usepackage{graphicx}

\usepackage[colorlinks=true,linkcolor=blue,citecolor=blue,urlcolor=blue]{hyperref}

\usepackage{epstopdf}
\usepackage{amssymb}
\usepackage{color}
\usepackage{url}
\usepackage[justification=justified]{subcaption}

\usepackage{float}
\restylefloat*{figure}
\usepackage{placeins}

\usepackage{dcolumn}
\usepackage{bm} 

\usepackage{siunitx}
\usepackage{textcomp} 
\usepackage{multirow}
\usepackage{tabularx} 
\usepackage{ragged2e}
\usepackage[overload]{textcase}

\newcommand{\isotope}[2]{$^{#2}$#1}

\begin{document}


\title{Cosmogenic production of $^{37}$Ar in the context of the LUX-ZEPLIN experiment}


\author{J.~Aalbers}
\affiliation{SLAC National Accelerator Laboratory, Menlo Park, CA 94025-7015, USA}
\affiliation{Kavli Institute for Particle Astrophysics and Cosmology, Stanford University, Stanford, CA  94305-4085 USA}

\author{D.S.~Akerib}
\affiliation{SLAC National Accelerator Laboratory, Menlo Park, CA 94025-7015, USA}
\affiliation{Kavli Institute for Particle Astrophysics and Cosmology, Stanford University, Stanford, CA  94305-4085 USA}

\author{A.K.~Al Musalhi}
\affiliation{University of Oxford, Department of Physics, Oxford OX1 3RH, UK}

\author{F.~Alder}
\affiliation{University College London (UCL), Department of Physics and Astronomy, London WC1E 6BT, UK}

\author{S.K.~Alsum}
\affiliation{University of Wisconsin-Madison, Department of Physics, Madison, WI 53706-1390, USA}

\author{C.S.~Amarasinghe}
\affiliation{University of Michigan, Randall Laboratory of Physics, Ann Arbor, MI 48109-1040, USA}

\author{A.~Ames}
\affiliation{SLAC National Accelerator Laboratory, Menlo Park, CA 94025-7015, USA}
\affiliation{Kavli Institute for Particle Astrophysics and Cosmology, Stanford University, Stanford, CA  94305-4085 USA}

\author{T.J.~Anderson}
\affiliation{SLAC National Accelerator Laboratory, Menlo Park, CA 94025-7015, USA}
\affiliation{Kavli Institute for Particle Astrophysics and Cosmology, Stanford University, Stanford, CA  94305-4085 USA}


\author{N.~Angelides}
\affiliation{University College London (UCL), Department of Physics and Astronomy, London WC1E 6BT, UK}

\author{H.M.~Ara\'{u}jo}
\affiliation{Imperial College London, Physics Department, Blackett Laboratory, London SW7 2AZ, UK}

% ...

\collaboration{The LUX-ZEPLIN Collaboration}

\begin{abstract}
\noindent
We estimate the amount of $^{37}$Ar produced in natural xenon via cosmic-ray-induced spallation, an inevitable consequence of the transportation and storage of xenon on the Earth's surface.
We then calculate the resulting $^{37}$Ar concentration in a 10-tonne payload~(similar to that of the LUX-ZEPLIN experiment) assuming a representative schedule of xenon purification, storage and delivery to the underground facility.
Using the spallation model by Silberberg and Tsao, the sea-level production rate of $^{37}$Ar in natural xenon is estimated to be 0.024~atoms/kg/day.
Assuming the xenon is successively purified to remove radioactive contaminants in 1-tonne batches at a rate of 1~tonne/month, the average $^{37}$Ar activity after 10~tonnes are purified and transported underground is 0.058--0.090~$\mu$Bq/kg, depending on the degree of argon removal during above-ground purification.
Such cosmogenic $^{37}$Ar will appear as a noticeable background in the early science data, while decaying with a 35-day half-life.
This newly noticed production mechanism of $^{37}$Ar should be considered when planning for future liquid-xenon-based experiments.

\end{abstract}


\maketitle

\section{Introduction}
\label{sec:intro}

Liquid xenon (LXe) time projection chambers~(TPCs) are the most sensitive technology searching for weakly interacting massive particle~(WIMP) dark matter via characteristic keV-scale nuclear recoils (NRs)~\cite{LUX:2016ggv,XENON:2018voc,PandaX-4T:2021bab}. In addition, these detectors are sensitive to numerous novel physics processes in the electron recoil (ER) channel~\cite{xenon1t-excess,lz-lowER-sensitivity}. To maximize their experimental sensitivity for rare processes, care must be taken to minimize backgrounds caused by cosmic rays, ambient gamma rays and neutrons, and radioactive isotopes within the LXe target itself. One potential source of background is the radioactive noble gas \isotope{Ar}{37}, which can contaminate the few-keV energy region where LXe TPCs are most sensitive to WIMP dark matter.
\isotope{Ar}{37} can be introduced into LXe as residuals of argon imuprities, via ambient air leaks and activation.

In this manuscript, we first describe the \isotope{Ar}{37} decay and its relevance to these searches.
We then discuss the cosmogenic production of \isotope{Ar}{37} in xenon and estimate its activity in the context of the LUX-ZEPLIN (LZ) experiment~\cite{lz-detector} assuming a simplified schedule of xenon purification, storage on the surface and delivery.
Finally, the impact on LZ backgrounds and physics searches is discussed.


\section{Experimental signature of \isotope{A\lowercase{r}}{37} in LX\lowercase{e} TPCs}
\label{sec:Ar37}
\noindent

The isotope \isotope{Ar}{37} decays to the ground state of \isotope{Cl}{37} by electron capture with a half-life of 35.01(2)~days~\cite{Cameron:2012ogv}. The subsequent atomic relaxation of the \isotope{Cl}{37} daughter results in energy deposits at the atomic scale:  K-shell (2.82~keV, 90.2\%), L-shell (0.270~keV, 8.9\%), and M-shell (0.018~keV, 0.9\%).  The K-shell capture results in some mixture of emitted Auger electrons and x rays with energies that sum to 2.82~keV.

Particle interactions in the active region of a LXe TPC generate both a scintillation~(S1) and an ionization~(S2) signal, the ratio of which can be used to identify events as ERs or NRs.  The S1 and S2 response of LXe TPCs to \isotope{Ar}{37} decay, in particular the 2.82~keV K-shell feature, has been observed and characterized both in small surface installations~\cite{pixey, akimov, Baudis:2020} and in large underground installations (including LUX~\cite{balajthy_thesis, boulton_thesis} and XENON1T~\cite{xenon1t-excess}).  The Noble Element Simulation Technique (NEST)~\cite{NEST2011, NESTv2.2.1p1} is a response model which well describes S1 and S2 production for low-energy ER sources~\cite{xe127, tritium} including \isotope{Ar}{37}~\cite{Szydagis:2020isq, Baudis:2020}.
The S2/S1 signal from \isotope{Ar}{37} electron capture may be slightly affected by the atomic relaxation following the K-shell vacancy, but a recent measurement of \isotope{Xe}{127} electron capture indicates this should be a very small effect in \isotope{Ar}{37}~\cite{temples2021measurement} and thus this effect is not considered here.

\begin{figure}[tb!]
    \includegraphics[width=\columnwidth]{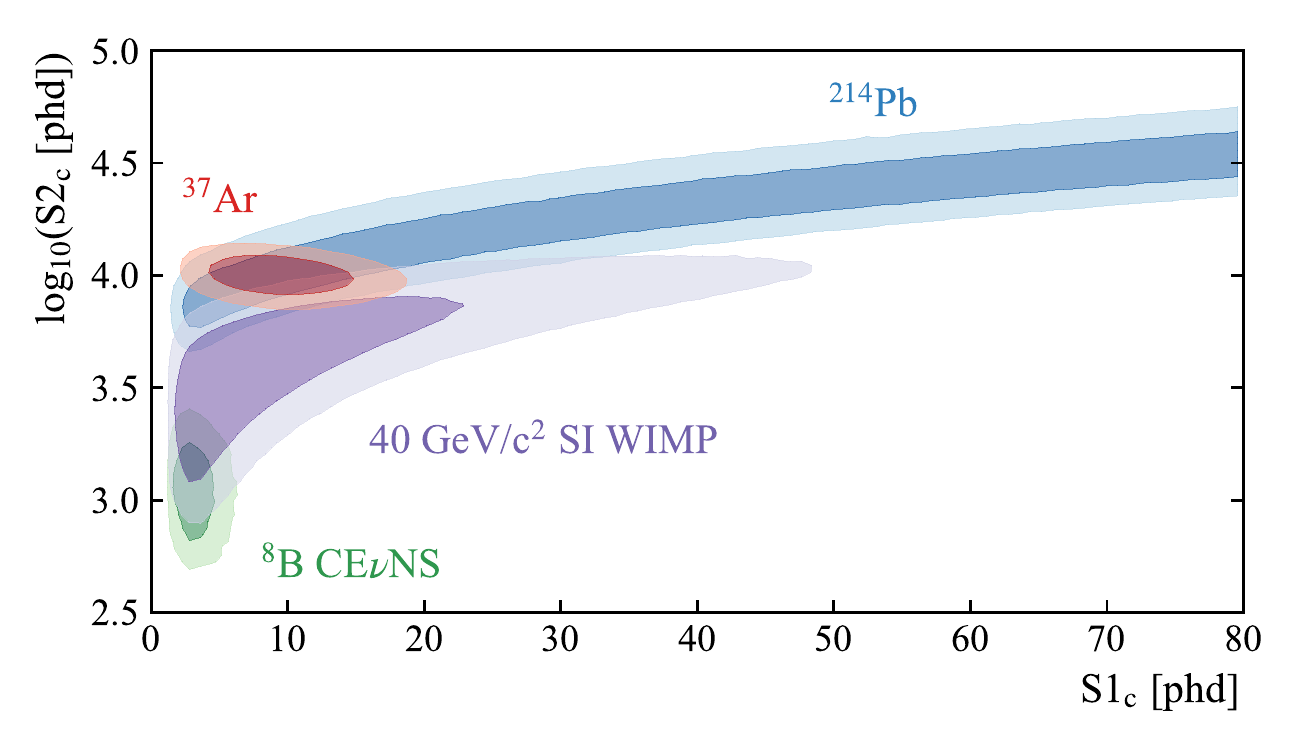}
    \caption{The distributions of \isotope{Ar}{37} decays and several other populations in the \{S1$_{\rm c}$, $\log_{10}$S2$_{\rm c}$\} plane (where S1$_{\rm c}$ and S2$_{\rm c}$ are S1 and S2 signals which have been corrected for position dependence within the TPC and phd denotes the number of photons detected) expected in LZ assuming the data selection described in Ref.~\cite{lz-wimp-sensitivity}.  Shown also are NRs from a 40~GeV/c$^2$ WIMP (purple), coherent elastic neutrino-nucleus scattering (${\rm CE}\nu{\rm NS}$) of \isotope{B}{8} solar neutrinos (green), and ground-state $\beta$ decays of \isotope{Pb}{214} (from dissolved \isotope{Rn}{222}) (blue). For each population, the dark and light regions indicate the $1\sigma$ and $2\sigma$ regions, respectively.}
    \label{fig:contour}
\end{figure}

Figure~\ref{fig:contour} shows the expected S1 vs log(S2) distribution of several populations in the LZ detector assuming the operating conditions and data selections described in Ref.~\cite{lz-wimp-sensitivity} and using the NESTv2.2.1patch1 model~\cite{NESTv2.2.1p1}. The $\beta$ decay of \isotope{Pb}{214} (a \isotope{Rn}{222} daughter) broadly populates the ER band, \isotope{B}{8} neutrinos produce NR signals at very low energies, and a typical 40 GeV/c$^2$ WIMP signature populates the NR region between the ER band and the \isotope{B}{8} neutrinos.  Also shown is the 2.82~keV K-shell decay of \isotope{Ar}{37}.
Its small but finite overlap with the WIMP distribution indicates that \isotope{Ar}{37} decay can weaken experimental sensitivity to a WIMP signal. More directly, this feature of \isotope{Ar}{37} forms a background in searches for novel physics processes at similar few-keV energies in the ER band, such as solar axion and neutrino magnetic moment interactions~\cite{lz-lowER-sensitivity}. The lower-energy L-shell and M-shell peaks may appear in analyses utilizing only the S2 signal, but they are typically below any anticipated S1 threshold.  Anticipating and modeling any potential \isotope{Ar}{37} background is particularly important given a recent observation from the XENON1T experiment of an excess of events in this low-energy ER region~\cite{xenon1t-excess,Szydagis:2020isq}.

\section{Cosmogenic production of \isotope{A\lowercase{r}}{37} }
\label{sec:production}

Argon-37 is found in small quantities in the atmosphere.  This \isotope{Ar}{37} can be generated by cosmic bombardment of atmospheric Ar, mostly via the spallation process \isotope{Ar}{40}$(\text{n},4\text{n})$\isotope{Ar}{37} but also via neutron capture on \isotope{Ar}{36}~\cite{ar37atmosphere, ar37atmosphere2}.  Atmospheric \isotope{Ar}{37} can also be produced by cosmic bombardment of calcium-containing soils, via \isotope{Ca}{40}$(\text{n},\alpha)$\isotope{Ar}{37}~\cite{ar37atmosphere3}. This atmospheric \isotope{Ar}{37} has been considered as a potential source of low-energy excess above other backgrounds by both the LUX experiment and the XENON1T experiment in the context of potential air leaks and residuals of initial argon contamination.~\cite{lux-data,xenon1t-excess}.

A separate production mechanism has not been previously considered in the literature: the cosmogenic production of \isotope{Ar}{37} in xenon itself via spallation of Xe by protons and neutrons~(more precisely, nuclear fragmentation).
This process has a nonzero cross section since spallation product yields are generally continuous in mass/atomic number, provided basic conservation laws are not violated~\cite{Russell:1990bq}.
Due to the large mass difference between Xe and \isotope{Ar}{37}, the production of \isotope{Ar}{37} from natural xenon by spallation is limited in rate and has not yet been observed experimentally.
The energy-dependent proton-induced spallation cross sections are frequently modeled using the semiempirical formula by Silberberg and Tsao~\cite{silberberg1973partial,Silberberg:1973partial2,silberberg1990spallation,Silberberg:1998}. In this model, the spallation cross section takes the form~\cite{silberberg1973partial}
\begin{eqnarray}
\label{eq:ST}
    \sigma &=& \sigma_0 \, \Omega \, \eta \, \xi \, f_{(A)}f_{(E)} \; \; e^{-P\;\Delta A} \;\;  e^{-R|Z-SA+TA^2|^\nu} \, , 
\end{eqnarray}
where $E$ is the incident proton energy, $A$ and $Z$ are the atomic mass and atomic number of the product nucleus, and $P$, $R$, $S$, $T$, and $\nu$ are empirical parameters.
The generic cross section behavior is captured in $\sigma_0$, which depends on the mass/atomic number of the product and target and also the incident proton energy. The functions $f_{(A)}$ and $f_{(E)}$ provide corrections when the product nucleus is produced from heavy targets and when the change in mass number~($\Delta A=A_t-A$) is large, respectively.
The first exponential term describes the decrease in cross section as the target-product mass difference becomes large, and the second exponential term describes the statistical distribution of various isotopes for a product of a given $Z$.
The three factors $\Omega$, $\eta$ and $\xi$ account for corrections due to nuclear structure, nuclear pairing, and enhancement of light evaporation products, respectively~\cite{silberberg1973partial}.
The  model's prediction is generally accurate to within a factor of 2 or 3, as assessed by comparing the predicted and experimentally measured cross sections for various target-product pairs at discrete energies~\cite{silberberg1973partial}.
The actual computation of spallation cross sections is more involved as many of the above-mentioned parameters~($\sigma_0$, $P$, $R$, $S$, $T$, $\nu$) take different expressions depending on the mass numbers of the target and product, and the incident energy.
Interested readers are referred to the original article~\cite{silberberg1973partial} for a complete description of the model.

\begin{figure}[tb!]
    \centering
    \includegraphics[width=\columnwidth]{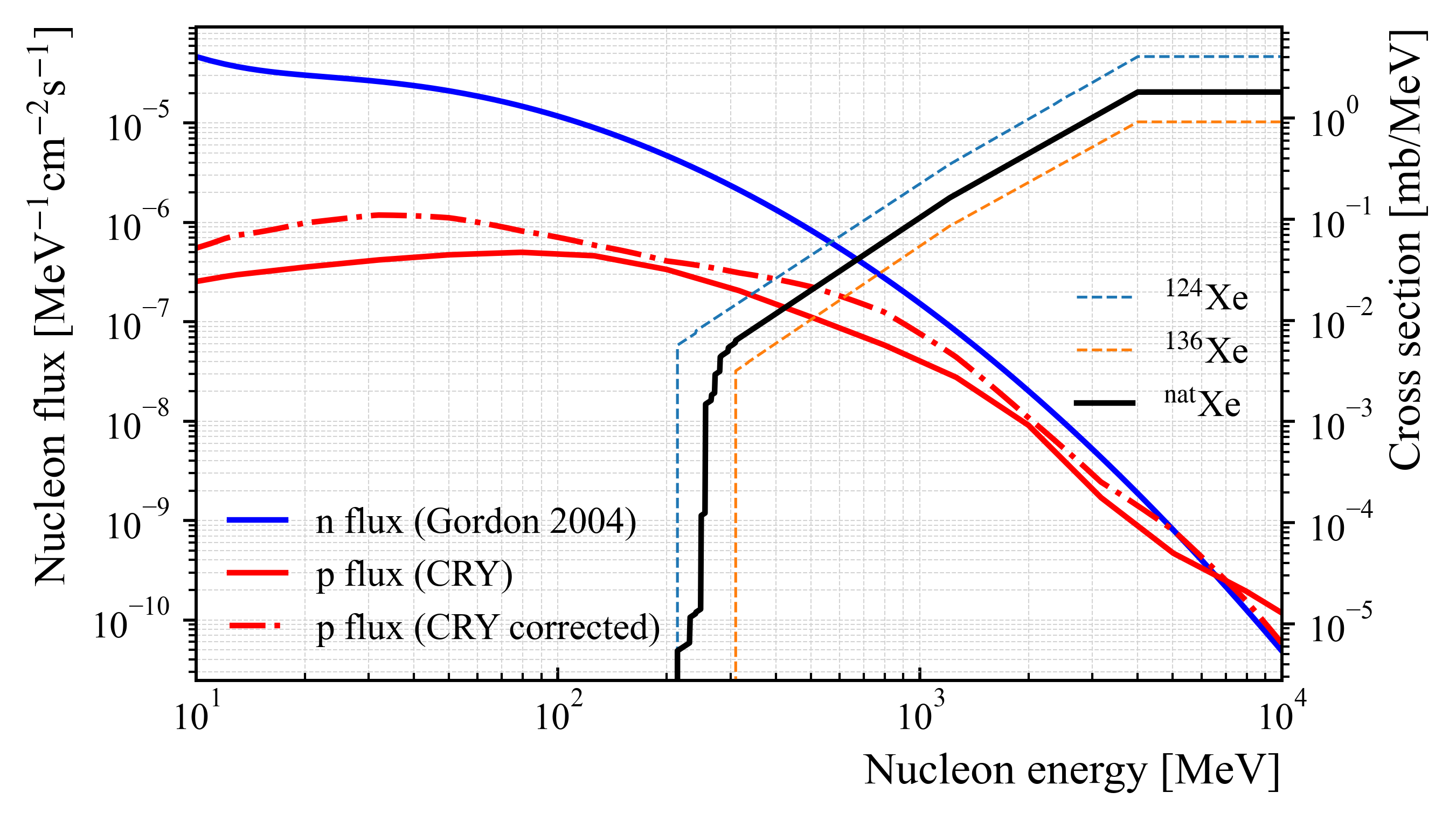}
    \caption{The calculated spallation cross section of \isotope{Ar}{37} from individual xenon isotopes (light dotted curves) and natural xenon (solid black). Overlaid is the surface nucleon flux used in our calculations~\cite{gordon2004neutron, cry2007cosmic}. According to the model of Silberberg and Tsao~\cite{silberberg1973partial}, the spallation cross section is negligible below about \SI{210}{\MeV} and increases with energy until \SI{4}{\GeV}, beyond which it is assumed to be constant.}
    \label{fig:xsec}
\end{figure}

Although the original Silberberg and Tsao model is formulated for proton-induced spallation, isospin invariance allows the model to also describe neutron-induced spallation at the relevant (high) energies of 100s of MeV and higher, obtained by cosmic-ray-induced neutrons.
The model is conveniently implemented in the \textsc{ACTIVIA} package~\cite{activia2008} and is frequently used to calculate activation due to neutrons~\cite{Baudis:2015kqa,Cebrian:2017oft}.
Figure~\ref{fig:xsec} (right-side vertical scale) shows the differential cross section of \isotope{Ar}{37} production from natural xenon by spallation as a function of incident nucleon energy.
The low-energy cutoff at approximately \SI{250}{\MeV} reflects the energy required by the incident nucleon to initiate an intranuclear cascade in the target nucleus~\cite{Serber:1947zza}.
Only the cross sections of the lightest and heaviest stable xenon isotopes are shown for clarity: all other stable isotopes lie between these two curves.  The black curve represents an average cross section, weighted by natural isotopic abundance.
In calculating the final production rate, the cosmic neutron energy spectrum measured by Gordon~\textit{et}~al.~\cite{gordon2004neutron} and the proton spectrum from the Cosmic-Ray Shower Generator~(CRY,  version~1.7)~\cite{cry2007cosmic} are used.
Since CRY accounts for products from protons in the primary cosmic ray only and hence underestimates the flux, the CRY proton spectrum is further scaled by the ratio of Gordon's neutron spectrum to CRY's neutron spectrum.  
These spectra are shown in Fig.~\ref{fig:xsec} (left-side vertical scale). 
The proton spectrum is generated at the latitude of New York City to be consistent with Gordon's measurement of the neutron spectrum~\cite{gordon2004neutron}.
A correction due to geomagnetic latitude is not included in the nucleon spectrum, as the geomagnetic rigidity cutoff in the locations of relevance in North America does not vary significantly enough compared to uncertainties due to other sources.
Temporal change in the nucleon flux is similarly not considered here.
The additional shielding due to building structure and storage material is not considered either.

\begin{figure}[tb!]
    \centering
    \includegraphics[width=\columnwidth]{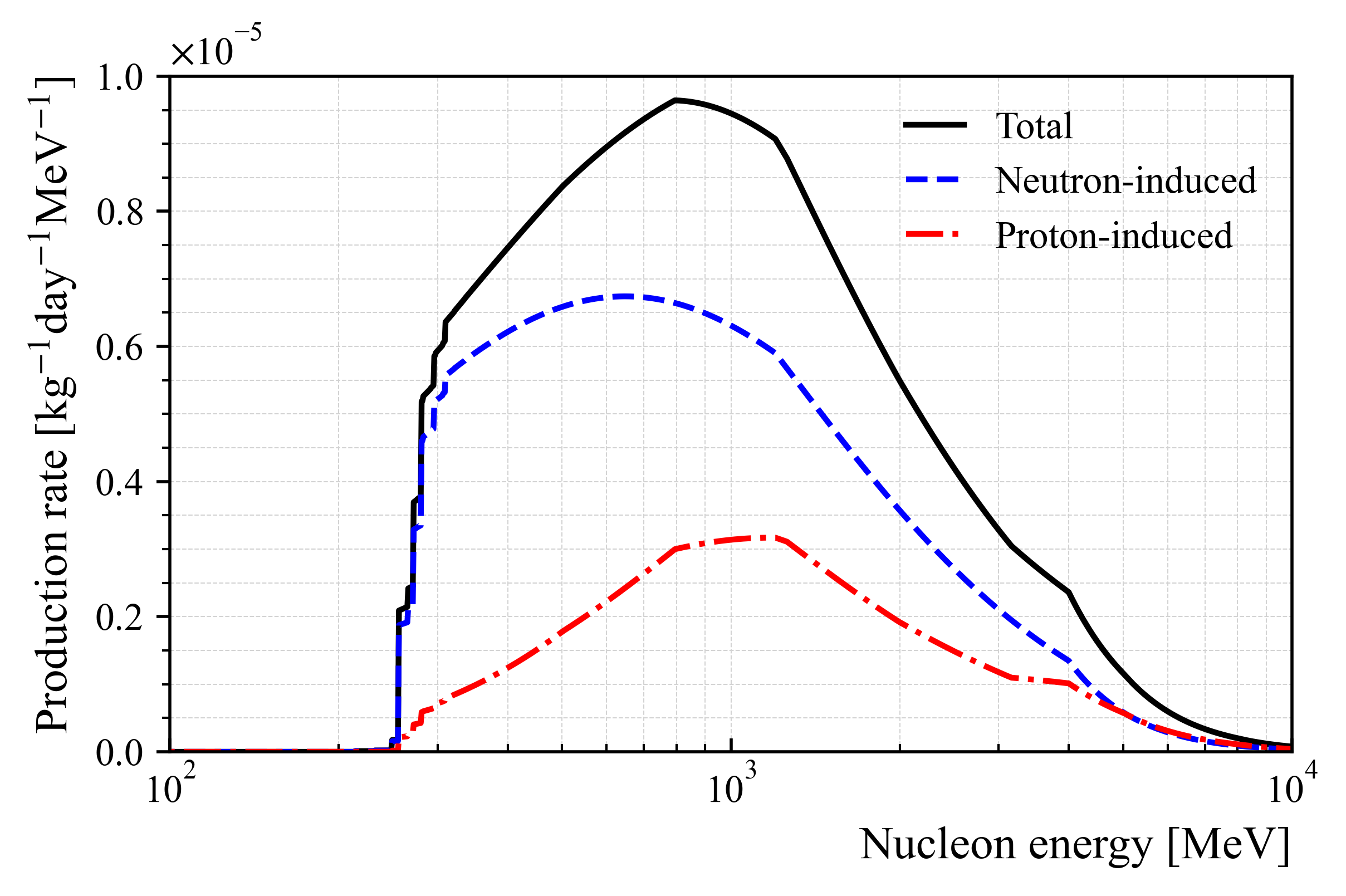}
    \caption{The differential production rate of \isotope{Ar}{37} in natural xenon via spallation as a function of the incident nucleon energy. The production is dominated by nucleons with energies between \SI{300}{\MeV} and \SI{2}{\GeV}. The decrease of production rate below \SI{300}{\MeV} is due to the nucleon energy being too low to initiate an intranuclear cascade, while on the higher end, the decrease of production rate is caused by the decrease of nucleon flux.}
    \label{fig:prodrate}
\end{figure}

As shown in Fig.~\ref{fig:xsec}, the spallation cross section increases towards higher incident nucleon energy whereas the cosmic proton and neutron fluxes decrease rapidly with energy.
As a result, \isotope{Ar}{37} production at sea level is dominated by protons and neutrons with energies between \SI{300}{\MeV} and a few GeV. The differential production rate of \isotope{Ar}{37} in natural xenon is shown in Fig.~\ref{fig:prodrate} as a function of nucleon energy. Upon integrating the differential rate, the final production rate of \isotope{Ar}{37} due to cosmogenic activation of natural xenon at sea-level is estimated to be 0.024~atoms/kg/day, subject to the same factor of 2 or 3 theoretical uncertainty pointed out earlier for the Silberberg and Tsao spallation model more generally.

Currently, there is no experimental data on the \isotope{Ar}{37} production cross section due to its relatively short half-life.
Although a partial measurement is possible in a neutron beam facility such as LANSCE~\cite{lansce,ar37atmosphere2}, due to the deviation of the neutron beam profile above 500~MeV from true cosmic neutrons and the increase of the production cross section towards higher energies, the calculation of the total production rate is still model dependent.
Therefore, we expect that an \emph{in situ} measurement of its concentration in LZ can provide data on the total, flux-weighted cross section.

The cosmogenic production of \isotope{Ar}{37} in Xe via spallation should be very limited in a deep underground setting.
The hadronic components of the cosmic rays are strongly attenuated by the rock overburden while the low-energy neutrons from spontaneous fission and $(\alpha,\text{n})$ reactions are below the spallation threshold energy.
Instead, the production of \isotope{Ar}{37} in xenon underground is dominated by muon-induced neutrons, of which the flux in the relevant energy range in a typical underground laboratory is $10^5$--$10^7$ times smaller than that on the surface~\cite{gordon2004neutron,mei2006muon}.

\section{Cosmogenic production of \isotope{A\lowercase{r}}{37} in the LZ context}
\label{sec:LZ}

The xenon used in the LZ experiment is purified to remove the radioisotope \isotope{Kr}{85}. This purification proceeds via gas-phase chromatography in charcoal at SLAC National Accelerator Laboratory~(California, USA)~\cite{LUX:2016wel}, which removes noble gas elements other than xenon. Although a detailed analysis is still in progress, preliminary data indicates that the argon concentration is reduced by at least a factor of 100 by the charcoal chromatography. As a result, the \isotope{Ar}{37} produced prior to chromatography is strongly suppressed.

After purification, the xenon is transported by road to the Sanford Underground Research Facility (SURF) in South Dakota, USA~\cite{lesko2015sanford} and brought underground to the LZ experiment site at a depth of \SI{-1480}{\meter}~(4300~m.w.e.).
Because of the argon removal during purification, the majority of \isotope{Ar}{37} activity is produced during storage and shipment (between purification at SLAC and delivery underground).
As will be shown later, the rate of cosmogenic production is rapid enough such that the argon reduction by chromatography does not play an important role.
During ground transportation to SURF, the production rate is also accelerated by the increased proton and neutron flux at higher altitudes, since the SURF surface facility is located at an altitude of 1600~m.
Once the xenon is brought underground, the production of \isotope{Ar}{37} in natural xenon becomes negligible, and the \isotope{Ar}{37} accumulated during the transportation decays exponentially over time.

As an illustrative model of this process, we assume a simplified schedule of xenon purification, storage and delivery, broadly representative of the actual xenon logistics in LZ.
We assume xenon is purified at SLAC in successive, 1-tonne batches at a rate of one batch/month, and we assume ten equal batches totaling 10 tonnes of xenon.
The batches are shipped from SLAC to SURF in pairs by ground transportation once every two months, and during each shipment it is assumed that the altitude increases linearly from 86~m above sea level~(at SLAC) to 1600~m~(at the SURF surface facility) over a three-day period. Once at SURF, we assume the xenon is immediately moved underground.
The incident proton and neutron flux is assumed to increase exponentially with altitude with attenuation coefficients of 110~g/cm$^2$ and 148~g/cm$^2$, respectively~\cite{ziegler1996}:
\begin{eqnarray}
    I_j = I_i e^{(A_i-A_j)/L} 
\end{eqnarray}
where $I_i$ and $I_j$ are the intensities at altitude $i$ and $j$, and $A_i$ and $A_j$ are the atmospheric depths of the respective locations. $L$ is the attenuation coefficient of the particle of concern. The atmospheric depth is defined as the integral of air density with respect to depth measured from the upper atmosphere. In the lower atmosphere, its difference can be approximated simply as density times height difference, namely $A_i-A_j=\rho (h_j-h_i)$.
This correction is applied uniformly to the proton and neutron spectra since the energy-dependent attenuation coefficient does not vary significantly over the energy range of interest~\cite{cry2007cosmic,ziegler1996}.
The \isotope{Ar}{37} production rate at higher altitudes is obtained by scaling the surface production rate with the elevation-specific increase in nucleon flux.

\begin{figure}[t]
\includegraphics[width=\columnwidth]{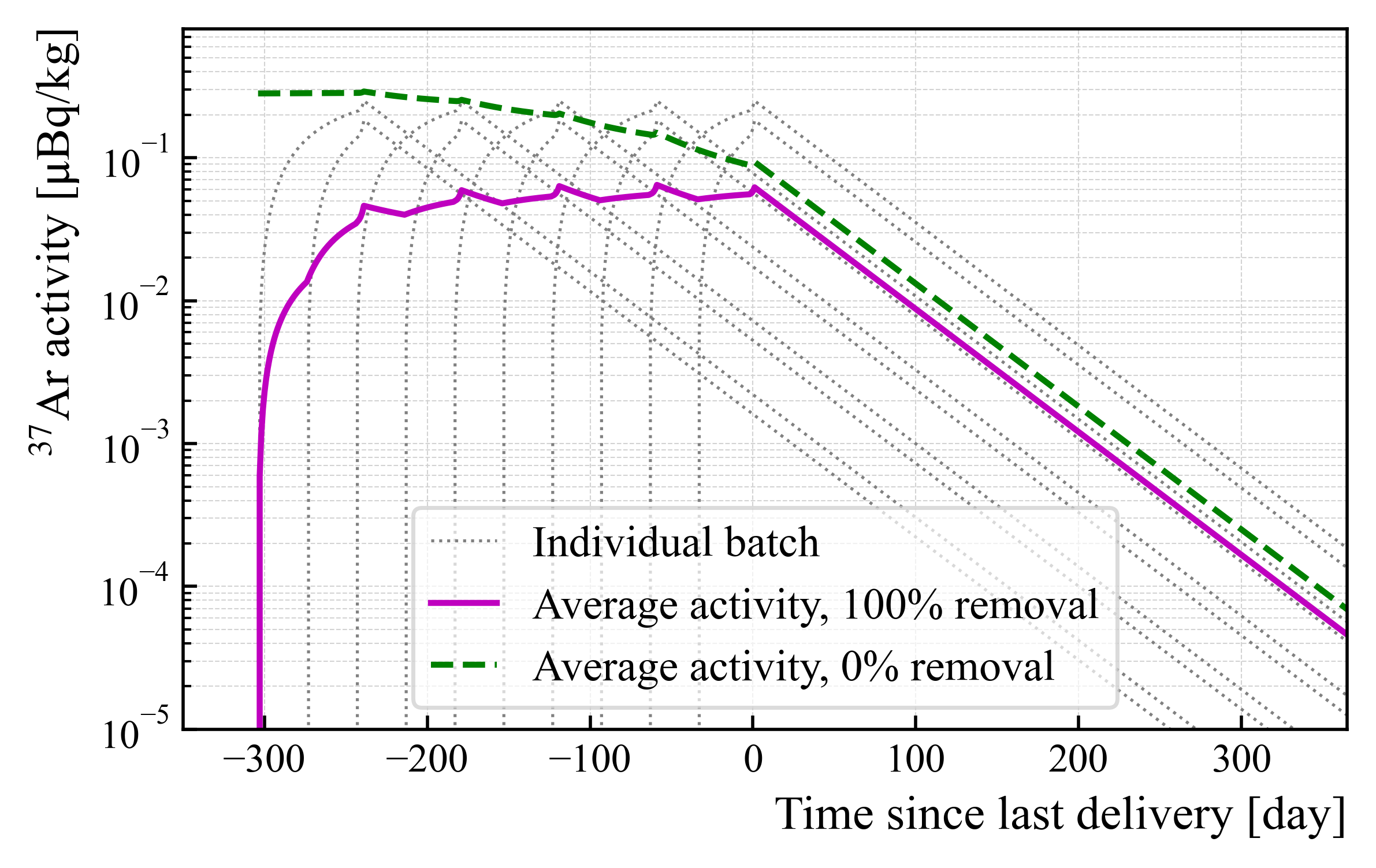}
\caption{\label{fig:ar37_transportation} 
Projected \isotope{Ar}{37} activity in the xenon following the simplified purification, storage, and transportation schedule described in the text (ten 1-tonne batches of xenon, delivered at monthly intervals, two batches per shipment). The dotted lines track the \isotope{Ar}{37} activity in each of the 1-tonne batches after they have undergone purification, assuming complete removal of argon is achieved.
Note that in each shipment group, as the second batch is being purified, the first batch is stored on the surface and \isotope{Ar}{37} continues to grow.
The solid magenta curve shows the average activity in the final 10-tonne payload under that same assumption.
The green dashed line shows the scenario when Ar is not removed during purification.}
\end{figure}

Figure~\ref{fig:ar37_transportation} shows the result of this simplified model of LZ logistics. The instantaneous \isotope{Ar}{37} activities in each 1~tonne batch are shown as faint dotted lines, beginning at the time of each batch's purification at SLAC.  Also shown (as a thick solid line) is the activity per unit mass in the purified xenon payload. Because the Ar removal efficiency of the chromatography at SLAC remains somewhat uncertain, we also show a conservative model in which chromatography results in no \isotope{Ar}{37} removal (dashed line).  Assuming complete removal of argon by purification at SLAC, the estimated \isotope{Ar}{37} activity at the time of last delivery is \SI{0.058}{\micro\becquerel/\kilogram}. If no argon is removed, the estimated activity on that date is roughly 50\% higher~(\SI{0.090}{\micro\becquerel/\kilogram}).  After this date of last delivery underground, the average activity falls with the 35-day half-life.  Notice that details of the production and delivery schedule of the last few batches will have a dominant effect on the final total activity as compared to the earlier batches.

The trace natural argon left in the xenon after purification can also be activated during storage and shipment to produce some amount of \isotope{Ar}{37}.  This cosmogenic production rate of \isotope{Ar}{37} in argon is about 5000~times higher than the rate of cosmogenic production of \isotope{Ar}{37} in natural xenon~\cite{ar37atmosphere3}.  However, taking the most extreme assumptions~(that argon is the only impurity in the initial 99.999\%-purity xenon\footnote{The actual concentration of argon in the xenon prior to chromatography is less than 30~ppb.}, and that the argon is not removed at any level during purification), we find that \isotope{Ar}{37} produced by activation of argon will be subdominant, accounting for at most 5\% of the total cosmogenic \isotope{Ar}{37} in LZ.

Argon-37 can also be produced in the plumbing and storage material---most notably steel---and subsequently diffuse into the xenon. The production rate of \isotope{Ar}{37} in iron by spallation is predicted to be around 2.4~atoms/kg/day by ACTIVIA. However, its contribution to radioactivity in the xenon is strongly limited by the slow diffusion rate of argon in steel: even if argon had the same diffusivity in steel as helium~(about $10^{-13}$~cm$^2$/s in common metals~\cite{helium-diffusion}), only a surface depth of a fraction of a millimeter can contribute to the xenon radioactivity over the timescale of a few months. In practice, argon diffusion is significantly slower than that of helium; thus, the contribution of \isotope{Ar}{37} produced in the steel housing material is negligible compared to \isotope{Ar}{37} produced in the bulk xenon.
Furthermore, the \isotope{Ar}{37} in the underground plumbing and storage material produced during surface exposures is negligible since these components have been underground for longer than the half-life of \isotope{Ar}{37}.
\emph{In situ} production of \isotope{Ar}{37} by spallation in these peripheral detector components is also suppressed by the number of neutrons with sufficient energy.
An exception occurs when the target mass number is close to that of Ar and the nuclear transmutation can be triggered by low-energy neutrons~(e.g. neutron capture by \isotope{Ca}{40}). Material contamination at these atomic masses should be given particular attention in future experiments.


\section{Impact on LZ Backgrounds and Physics Searches}\label{sec:impact}

\begin{figure}[tb]
    \includegraphics[width=\columnwidth]{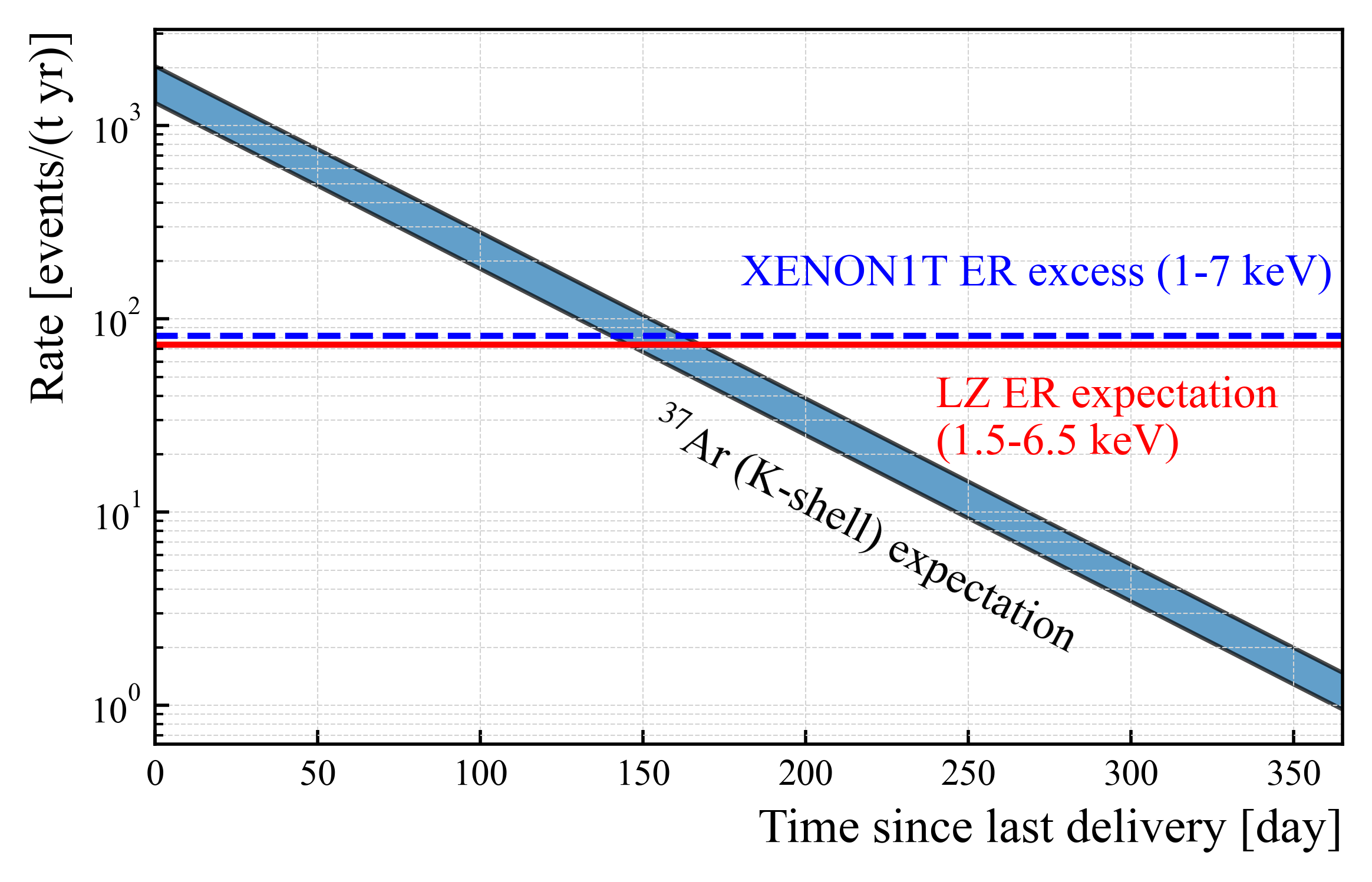}
    \caption{Expected event rate from cosmogenic \isotope{Ar}{37} since the time the last batch of xenon is delivered underground. The width of the band indicates variation from assuming either complete or negligible Ar removal during purification at SLAC. The blue dashed line shows the rate of excess observed in this region above the best-fit background model in the XENON1T experiment~\cite{xenon1t-excess}. The solid red line shows the rate of expected ER backgrounds in the LZ experiment, integrated over a 1.5--6.5~keV window relevant for a 40~GeV/c$^2$ WIMP and several ER new-physics signals~\cite{lz-wimp-sensitivity,lz-lowER-sensitivity}.
    }
    \label{fig:activity_vs_time}
\end{figure}

Figure~\ref{fig:activity_vs_time} shows the time evolution of the \isotope{Ar}{37} event rate~(after the same selection criteria as Fig.~\ref{fig:contour}) after the last xenon batch arrives underground assuming the delivery schedule of the previous section.
The width of the band represents the assumptions of either perfect or negligible Ar removal during gas chromatography.
The band does not include the uncertainties in the spallation cross section estimate from Silberberg's model.
For comparison, two other activities are shown: the expected rate of other LZ backgrounds in the ER band in a 1.5--6.5~keV window~(predominantly the \isotope{Pb}{214} daughter of \isotope{Rn}{222})~\cite{lz-wimp-sensitivity} and the rate of the excess seen in the 1--7~keV window by XENON1T in Ref.~\cite{xenon1t-excess}.
Initially, the \isotope{Ar}{37} K-shell feature is seen to be a dominant background in this window, weakening early sensitivity to novel physics processes via ERs.
\isotope{Ar}{37} becomes subdominant as it decays: at about 150~days since last delivery, the \isotope{Ar}{37} event rate is comparable to both the XENON1T excess rate and other ER background rates in the LZ detector.
After this point, the detector begins to reach its optimal sensitivity to the ER excess signal seen by the XENON1T or other novel physics processes in the low-energy ER channel.

Previous work has quantified the effect of an unknown constant \isotope{Ar}{37} rate in limiting LZ's sensitivity to several specific ER signals~\cite{lz-lowER-sensitivity}.  However, if the predominant source of \isotope{Ar}{37} is steadily decaying over a 1000~day run, then the mean \isotope{Ar}{37} activity is 20~times smaller than the instantaneous activity at the beginning of the run.
This aids in the statistical inference for new physics in the few-keV region since a fit to the \isotope{Ar}{37} rate early in the exposure reduces the rate uncertainty for later times in the run.

\section{Conclusions}

The noble radioisotope \isotope{Ar}{37} is a background of concern for LXe-based detectors searching for new physics at the few-keV energy scale.
Estimations of the production rate of \isotope{Ar}{37} in natural xenon via cosmic-ray-induced spallation yield \SI{0.024}{atoms\per\kilogram\per day} at sea level, subject to a model uncertainty of a factor of 2 or 3. Using a simplified model of the LZ xenon purification, storage and transportation schedule, the \isotope{Ar}{37} activity in the LZ payload is estimated to be 0.058--\SI{0.090}{\micro\becquerel/\kilogram} on the date when the last xenon is delivered underground. The upper (lower) bound assumes no removal (complete removal) of argon during the above-ground purification process. This is an experimental uncertainty which does not include the uncertainty in the spallation cross section estimated using Silberberg and Tsao's model.

The K-shell electron capture  of \isotope{Ar}{37} will likely appear as a significant background feature at 2.82~keV in early LZ data, due to the large quantity of recently above-ground xenon and the expected exceptionally low rate of all other backgrounds. This background will gradually become subdominant compared to other ER backgrounds (primarily \isotope{Pb}{214}) as it decays with a 35-day half-life.
The statistical strength of long-duration searches can be increased by taking advantage of the time dependence in this background component over the course of the exposure.

While the \isotope{Ar}{37} background has only a minimal effect on the primary physics goals of LZ, the effect can potentially be greater in future LXe experiments with increased target masses and decreased backgrounds. 
The cosmogenic production of \isotope{Ar}{37} in natural xenon via spallation discussed here should therefore be considered when planning these future experiments.
The timing of xenon handling and purification activities above ground should be optimized to limit \isotope{Ar}{37} activity in the purified xenon brought underground.  Ideally, xenon would be stored underground as early as possible in the logistics chain after purification.
The present work also highlights how the capacity to separate noble elements in the underground environment is important for future experiments.
The XMASS, XENON1T and XENONnT experiments have demonstrated a system of underground cryogenic distillation to this effect~\cite{xmass-distillation,xenon-distillation}, followed by the PandaX Collaboration~\cite{distillation_pandax}.  This cryogenic distillation method or some similar method (e.g., membrane methods~\cite{Kr_removal_membrane}) for underground removal of \isotope{Ar}{37} should now be considered an essential element in the design of future experiments.





\begin{acknowledgments}
The research supporting this work took place in whole or in part at the Sanford Underground Research Facility (SURF) in Lead, South Dakota. Funding for this work is supported by the U.S. Department of Energy, Office of Science, Office of High Energy Physics under Contracts Number DE-AC02-05CH11231, DE-SC0020216, DE-SC0012704, DE-SC0010010, DE-AC02-07CH11359, DE-SC0012161, DE-SC0014223, DE-SC0010813, DE-SC0009999, DE-NA0003180, DE-SC0011702, DE-SC0010072, DE-SC0015708, DE-SC0006605, DE-SC0008475, DE-FG02-10ER46709, UW PRJ82AJ, DE-SC0013542, DE-AC02-76SF00515, DE-SC0018982, DE-SC0019066, DE-SC0015535, DE-SC0019193 DE-AC52-07NA27344, and DOE-SC0012447. This research was also supported by U.S. National Science Foundation (NSF); the U.K. Science \& Technology Facilities Council under Grants number ST/M003655/1, ST/M003981/1, ST/M003744/1, ST/M003639/1, ST/M003604/1, ST/R003181/1, ST/M003469/1, ST/S000739/1, ST/S000666/1, ST/S000828/1, ST/S000879/1, ST/S000933/1, ST/S000747/1, ST/S000801/1, and ST/R003181/1 (JD); Portuguese Foundation for Science and Technology (FCT) under Grants number PTDC/FIS-PAR/2831/2020; the Institute for Basic Science, Korea (budget numbers IBS-R016-D1). We acknowledge additional support from the STFC Boulby Underground Laboratory in the U.K., the GridPP and IRIS Consortium, in particular at Imperial College London and additional support by the University College London (UCL) Cosmoparticle Initiative. This research used resources of the National Energy Research Scientific Computing Center, a DOE Office of Science User Facility supported by the Office of Science of the U.S. Department of Energy under Contract No. DE-AC02-05CH11231. This work was completed in part with resources provided by the University of Massachusetts' Green High Performance Computing Cluster (GHPCC). The University of Edinburgh is a charitable body, registered in Scotland, with the registration number SC005336. The assistance of SURF and its personnel in providing physical access and general logistical and technical support is acknowledged.

\end{acknowledgments}

\newpage

\bibliographystyle{apsrev4-2}
\bibliography{reference}

\end{document}